\documentclass[5p,times,twocolumn]{elsarticle}

\usepackage{amsmath,amssymb}
\biboptions{comma,sort&compress}
\usepackage[utf8]{inputenc}
\usepackage{graphicx}
\usepackage{bm}
\usepackage{hyperref}
\hypersetup{
colorlinks = true,
linkcolor = blue,
anchorcolor = blue,
citecolor = blue,
filecolor = blue,
urlcolor = blue
}
\usepackage[dvipsnames]{xcolor}

\graphicspath{{figs/}}

\journal{Physics Letters B}
\begin{document}
\begin{frontmatter}

\title{Kinematic and dynamical origins of mean-$p_T$ fluctuations in heavy-ion collisions}

\author[first,second]{Lipei Du}
\ead{ldu2@lbl.gov}
\affiliation[first]{organization={Department of Physics, University of California},
            city={Berkeley},
            postcode={94270}, 
            state={CA},
            country={USA}}
\affiliation[second]{organization={Nuclear Science Division, Lawrence Berkeley National Laboratory},
            city={Berkeley},
            postcode={94270}, 
            state={CA},
            country={USA}}
\date{\today}

\begin{abstract}
Event-by-event fluctuations of the mean transverse momentum (mean-$p_T$) provide a
sensitive probe of collective dynamics beyond single-particle spectra
and anisotropic flow.
We present a systematic study of mean-$p_T$ fluctuation observables using a Bayesian-calibrated multistage hydrodynamic framework, including quantitative comparisons to RHIC measurements and model-based investigations of beam-energy and kinematic-acceptance effects.
The experimental definitions employed by the STAR and ALICE
Collaborations are implemented explicitly and found to yield consistent results within controlled limits.
We study the centrality and beam-energy dependence of the observable,
its sensitivity to key soft-sector ingredients, and the impact of the
kinematic $p_T$ acceptance.
By introducing scaled-$p_T$ cuts, we demonstrate that a part of the apparent energy dependence arises from kinematic projection
effects, while the remaining trends reflect genuine collective dynamics.
Our results establish mean-$p_T$ fluctuations as a nontrivial and
independent validation of calibrated hydrodynamic descriptions of the
quark--gluon plasma.
\end{abstract}


\end{frontmatter}

\section{Introduction}\label{sec:intro}

Relativistic heavy-ion collisions produce strongly interacting matter
that exhibits pronounced collective behavior, commonly described as the
hydrodynamic expansion of a quark--gluon plasma (QGP)
\cite{Ollitrault:1992bk,Heinz:2013th}.
Extensive experimental and theoretical studies have shown that
single-particle spectra and anisotropic flow coefficients provide
stringent constraints on the bulk properties and transport coefficients
of the medium
\cite{Voloshin:2008dg,Song:2010mg,Bernhard:2019bmu,Nijs:2020ors,Nijs:2020roc,JETSCAPE:2020shq,JETSCAPE:2020mzn}.
Event-by-event fluctuations offer complementary information by probing
the response of the system to initial-state inhomogeneities and
dynamical fluctuations that are not accessible through event-averaged
observables
\cite{Jeon:2003gk,Gavin:2006xd}.

Fluctuations of the event-wise mean transverse momentum are among the most interesting
event-by-event observables.
Experimental measurements have revealed
characteristic centrality and beam-energy systematics, including an
overall dilution of fluctuations with increasing system size and a
nontrivial evolution from RHIC to LHC energies
\cite{CERES:2003sap,STAR:2005vxr,STAR:2019dow,ALICE:2014gvd,ALICE:2024apz,ATLAS:2024jvf}.
These trends have been discussed in terms of partial thermalization,
the number of effective particle-emitting sources, and the growth of
collective radial expansion
\cite{Gavin:2003cb,Gavin:2006xd,Gavin:2008ev,Bozek:2016yoj}.
Despite this long history, a quantitative interpretation of mean-$p_T$
fluctuations within a modern, globally calibrated multistage hydrodynamic framework
has remained limited.

Recent progress in Bayesian parameter estimation \cite{Paquet:2023rfd} has enabled the
systematic calibration of multistage hydrodynamic models to a broad set
of soft-sector observables, including particle yields, transverse-momentum
spectra, and anisotropic flow coefficients
\cite{Bernhard:2019bmu,Nijs:2020ors,Nijs:2020roc,JETSCAPE:2020shq,JETSCAPE:2020mzn}.
Such calibrations provide a controlled baseline for confronting
additional observables that were not included in the fitting procedure.
Mean-$p_T$ fluctuations are particularly well suited for this purpose,
as they are sensitive to event-by-event variations of radial flow,
freeze-out conditions, and late-stage hadronic dynamics, while being
only weakly constrained by the observables typically used in the
calibration.

An additional challenge in interpreting existing measurements is that
different experiments employ distinct definitions and weighting
procedures for mean-$p_T$ fluctuation observables.
In particular, the prescriptions used by the STAR and ALICE
Collaborations differ in their treatment of particle pairs,
multiplicity weighting, and kinematic acceptance
\cite{STAR:2019dow,ALICE:2014gvd,ALICE:2024apz}.
A meaningful model--data comparison therefore requires an explicit
implementation of the experimental definitions, rather than relying on
a single idealized observable.

In this Letter, we present a systematic study of mean-$p_T$ fluctuations using a Bayesian-calibrated multistage hydrodynamic model~\cite{JETSCAPE:2020shq,JETSCAPE:2020mzn}, including quantitative comparisons to RHIC measurements and predictive investigations of beam-energy and kinematic-acceptance effects at LHC energies.
We implement the STAR and ALICE definitions explicitly, examine the
centrality and beam-energy dependence of the observable, and study its
sensitivity to key soft-sector model ingredients.
We further investigate the role of the kinematic $p_T$ acceptance by
comparing results obtained with experimental and scaled $p_T$ cuts,
allowing us to disentangle genuine dynamical trends from kinematic
projection effects.

The remainder of this Letter is organized as follows.
Section~\ref{sec:model} describes the hydrodynamic framework and the
analysis setup, including the implementation of the experimental
definitions of the transverse-momentum fluctuation observable.
Section~\ref{sec:results} presents the main results, focusing on the
centrality and beam-energy dependence of mean-$p_T$ fluctuations, their
sensitivity to key model ingredients, and the role of kinematic
acceptance effects.
Finally, Section~\ref{sec:conclusions} summarizes the main findings and
outlines implications for future fluctuation measurements.

\section{Model and analysis setup}
\label{sec:model}

\subsection{Bayesian-calibrated multistage hydrodynamic framework}
The model calculations presented in this work are based on the same
Bayesian-calibrated multistage hydrodynamic framework \cite{JETSCAPE:2020shq,JETSCAPE:2020mzn,JETSCAPE:2022cob} employed in our
recent study of transverse-momentum–differential radial flow fluctuations,
$v_0(p_T)$~\cite{Du:2025dpu}.  Here we summarize only the essential ingredients
relevant for the present analysis of mean-$p_T$ fluctuations and refer
the reader to Refs.~\cite{JETSCAPE:2020shq,JETSCAPE:2020mzn,JETSCAPE:2022cob,Du:2025dpu} for a comprehensive description and
validation.

The simulations are performed within the JETSCAPE framework (v3.7)~\cite{Putschke:2019yrg} using a
TRENTo-based initial condition model~\cite{Moreland:2014oya}, followed by a
short period of pre-equilibrium free streaming~\cite{Broniowski:2008qk,Liu:2015nwa}
and subsequent second-order viscous hydrodynamic evolution with MUSIC~\cite{Schenke:2010nt,Schenke:2011bn,Paquet:2015lta}.
The hydrodynamic stage includes temperature-dependent shear and bulk viscosities
constrained by Bayesian inference to a broad set of soft-sector observables at both
RHIC and LHC energies~\cite{JETSCAPE:2020shq,JETSCAPE:2020mzn,JETSCAPE:2022cob}.
Hadronization is implemented via Cooper--Frye sampling at a switching temperature
$T_{\mathrm{sw}}$~\cite{Cooper:1974mv}, with viscous corrections to the distribution
function~\cite{Shen:2014vra,McNelis:2019auj}, and the subsequent hadronic rescattering
stage is modeled using the SMASH transport code~\cite{SMASH:2016zqf}.

In the present work, we use the Maximum A Posteriori (MAP) parameter set
from the Bayesian analysis~\cite{JETSCAPE:2020shq,JETSCAPE:2020mzn,JETSCAPE:2022cob} as a fixed baseline and perform
forward model calculations without retuning.  This setup provides a predictive test
of the calibrated framework at $\sqrt{s_{NN}}\!=\!200$~GeV and 5.02~TeV, since the
mean-$p_T$ fluctuation observables at these energies were not included in the
original calibration, while the corresponding data at $\sqrt{s_{NN}}\!=\!2.76$~TeV
were~\cite{JETSCAPE:2020shq,JETSCAPE:2020mzn}.  Variations around this baseline \cite{Du:2025dpu}—such
as switching off bulk viscosity, removing the hadronic afterburner, modifying the
nucleon size in the initial conditions, or considering ideal hydrodynamic
evolution—are used to assess the sensitivity of the observable to different stages
of the collision dynamics.

\subsection{Definition of the transverse-momentum fluctuation observable}
\label{subsec:observable_definition}
The observable studied in this work is an event-by-event measure of
correlated transverse-momentum fluctuations, designed to quantify
dynamical fluctuations beyond trivial statistical sampling.
Unlike single-particle spectra or average flow coefficients, this
observable probes correlations between particles within the same event
and is therefore sensitive to event-wise fluctuations of collective
radial expansion, freeze-out conditions, and late-stage dynamics.

At the core of the analysis is the two-particle transverse-momentum
covariance $C_m$, evaluated for particles within a given kinematic
acceptance and centrality class.
For a single event containing $N$ particles, the covariance is defined
as \cite{Voloshin:2002ku,Voloshin:2003ud}
\begin{equation}
C_m^{(\mathrm{event})}
=
\frac{1}{N(N-1)}
\sum_{i\neq j}
( p_{T,i} - p_{T,\,\mathrm{ref}} )
( p_{T,j} - p_{T,\,\mathrm{ref}} ),
\label{eq:Cm_event}
\end{equation}
where $p_{T,i}$ denotes the transverse momentum of particle $i$ and
$p_{T,\,\mathrm{ref}}$ is a reference mean transverse
momentum.
The precise definition of this reference mean distinguishes the
experimental prescriptions used by different collaborations, as
discussed below.

The experimentally reported quantity is obtained by averaging the
event-wise covariance over all events within a given centrality bin,
$C_m\!
=
\!\left\langle C_m^{(\mathrm{event})} \right\rangle_{\mathrm{events}}$,
and normalizing it by the squared mean transverse momentum.
Specifically, we define the dimensionless fluctuation measure
$$R_{p_T}
\equiv
\sqrt{C_m}/\langle p_T \rangle,$$
where $\langle p_T \rangle\!\equiv\!\langle [p_T] \rangle$ denotes the ensemble-averaged mean transverse
momentum in the same centrality class and acceptance.
This normalization removes trivial scaling with the absolute momentum
scale and enables meaningful comparisons across centrality, beam
energy, and particle species.\footnote{Residual kinematic effects associated with the finite $p_T$ acceptance can still influence such comparisons, as discussed in Sec.~\ref{sec:results}.}

Physically, $R_{p_T}$ quantifies the relative magnitude of correlated
momentum fluctuations within an event.
Event-by-event variations of the collective transverse velocity field
lead to coherent shifts of particle momenta, generating positive
two-particle covariances that survive the averaging over events.
As a result, $R_{p_T}$ is particularly sensitive to fluctuations of
radial collectivity and to mechanisms that amplify or damp such
fluctuations, including viscous effects, initial-state granularity, and
hadronic rescattering.

\subsection{Implementation of experimental definitions}
\label{subsec:experimental_definitions}

In this work, we explicitly implement the transverse-momentum
fluctuation definitions used by the STAR and ALICE Collaborations at the
particle level, applying identical kinematic cuts to the model and the data \cite{STAR:2019dow,ALICE:2014gvd,ALICE:2024apz}.
The two prescriptions differ in the choice of the reference mean
$p_{T,\,\mathrm{ref}}$ appearing in
Eq.~\eqref{eq:Cm_event}.
In the STAR definition, $p_{T,\,\mathrm{ref}}$ is taken to
be the ensemble-averaged mean transverse momentum within a given
centrality class,
$p_{T,\,\mathrm{ref}}^{\mathrm{STAR}}
=
\langle [p_T] \rangle_{\mathrm{events}}$,
where $[p_T]$ is the event-wise mean transverse momentum.
In contrast, the ALICE prescription uses the event-wise mean itself as
the reference,
$p_{T,\,\mathrm{ref}}^{\mathrm{ALICE}}
=
[p_T]^{(\mathrm{event})}$.

As shown analytically in \ref{app:method_comparison}, the STAR
observable can be decomposed into the ALICE-style two-particle covariance
plus an additional term proportional to the variance of the event-wise
mean transverse momentum, $\big\langle\big([p_T]_i-\langle p_T\rangle\big)^2\big\rangle$. The magnitude of this additional contribution
depends on particle multiplicity and the width of the centrality bin,
and may in principle lead to quantitative differences between the two
definitions.

\section{Results and discussion}
\label{sec:results}

\subsection{Baseline comparison and sensitivity to soft kinematics}
\label{subsec:baseline_200}
\begin{figure}[t]
    \centering 
    \includegraphics[width= 0.8\linewidth]{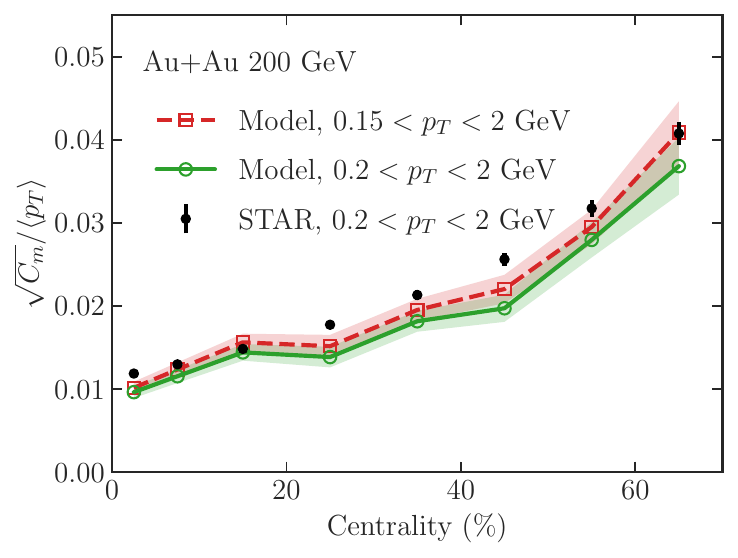}
    \caption{Centrality dependence of 
    $\sqrt{C_m}/\langle p_T\rangle$ in Au+Au collisions at
    $\sqrt{s_{NN}}=200$~GeV.
    Model calculations are shown for two different lower transverse-momentum
    cuts, $0.15<p_T<2$~GeV and $0.2<p_T<2$~GeV, illustrating the sensitivity
    of the integrated observable to the kinematic acceptance.
    Published STAR data for $0.2<p_T<2$~GeV are shown for comparison~\cite{STAR:2019dow}.}
    \label{fig:pt_fluctuation_ptcut}%
\end{figure}

We begin with a baseline comparison of $\sqrt{C_m}/\langle p_T\rangle$ at $\sqrt{s_{NN}}=200$~GeV, shown in Fig.~\ref{fig:pt_fluctuation_ptcut},
where model calculations are confronted with STAR measurements as a function of
centrality \cite{STAR:2019dow}. The results are evaluated using the experimental kinematic
acceptance $0.2<p_T<2$~GeV as in the data.
Applying both the STAR and ALICE fluctuation definitions to the same set
of model events with identical kinematic cuts, we find that the two
prescriptions yield numerically indistinguishable results across all
centralities considered. This confirms that, for the large multiplicities
and narrow centrality bins employed here, the additional variance term
discussed in \ref{app:method_comparison} is subleading. In view
of this near-equivalence, only the STAR-style result is shown in
Fig.~\ref{fig:pt_fluctuation_ptcut}.

While the STAR- and ALICE-style model evaluations are mutually consistent, the
model systematically underestimates the measured fluctuation magnitude when the
experimental lower $p_T$ cut of $0.2$~GeV is used. A similar underestimation is observed in the Bayesian inference study at
$\sqrt{s_{NN}}=2.76$~TeV, where mean-$p_T$ fluctuation data were included in the
calibration, yet the MAP parameter set nevertheless yields systematically smaller
fluctuation magnitudes for the experimental acceptance~\cite{JETSCAPE:2020mzn}. This indicates that the trend observed at 200 GeV reflects a systematic feature of the calibrated framework rather than an isolated discrepancy.\footnote{Mean-$p_T$ fluctuation measurements at LHC energies are available from ALICE and
are reported primarily as a function of charged-particle multiplicity rather
than in wide centrality classes \cite{ALICE:2014gvd,ALICE:2024apz}. Performing a direct model--data comparison
within the present framework would therefore require a dedicated reanalysis
using multiplicity-based event selection and narrow multiplicity bins, which in
turn would demand substantially higher event statistics to control statistical
fluctuations. Such an analysis is beyond the scope of the present work.
}
To investigate the origin of this difference, we
varied the lower edge of the acceptance window within the model while keeping the
upper cut fixed. As illustrated in Fig.~\ref{fig:pt_fluctuation_ptcut}, lowering
$p_T^{\min}$ leads to a noticeable increase of
$\sqrt{C_m}/\langle p_T\rangle$ and improves the agreement with the data.

This sensitivity to the lower $p_T$ cut indicates that the
$p_T$-integrated fluctuation observable is dominated by correlations carried by
the very soft part of the spectrum, where collective radial dynamics are most
pronounced. Schematically, the event-averaged
covariance can be written as
\begin{equation}
C_m \sim \int dp_T\,dp_T'\,
\frac{dN}{dp_T}\frac{dN}{dp_T'}\,
\langle \delta p_T\,\delta p_T' \rangle ,
\end{equation}
illustrating that the dominant contribution arises from the low-$p_T$ region
where particle yields ($dN/dp_T$) are largest and momentum shifts induced by collective
flow fluctuations are coherent (see also~\ref{app:kinematic_projection}). The residual discrepancy between the model and
the data for the nominal experimental cuts therefore suggests that, within the
present setup, the strength of event-by-event radial collectivity in the soft
sector may be slightly underestimated. Crucially, the near-identical results
obtained using the STAR and ALICE definitions demonstrate that this conclusion
is robust with respect to reasonable variations in the fluctuation definition
and is not an artifact of the averaging procedure.

These observations motivate the more systematic investigation of kinematic
acceptance effects presented in the following sections, where we examine how
the projection of momentum correlations onto different $p_T$ windows influences
comparisons across beam energies and particle species.

\subsection{Sensitivity to model ingredients at fixed beam energy}
\label{subsec:model_sensitivity}
\begin{figure}[t]
    \centering 
    \includegraphics[width= 0.8\linewidth]{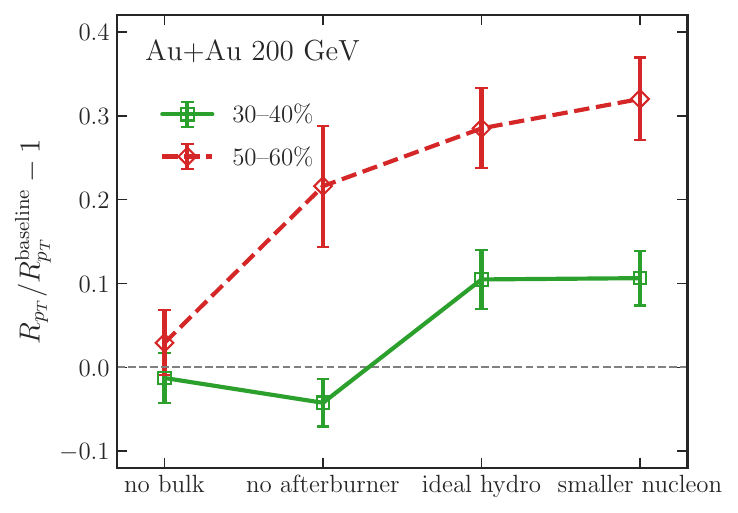}
    \caption{Relative response of 
    \(R_{p_T}\!\equiv\!\sqrt{C_m}/\langle p_T\rangle \) to variations of model ingredients
    in Au+Au collisions at \( \sqrt{s_{NN}} = 200 \) GeV. Results are shown as fractional
    deviations with respect to the baseline calculation that includes shear and bulk viscosity,
    finite nucleon size, and a hadronic afterburner. Two representative centrality
    classes, 30–40\% and 50–60\%, are shown. The model variants correspond to switching
    off bulk viscosity (``no bulk''), disabling the hadronic afterburner (``no
    afterburner''), ideal hydrodynamic evolution without viscosity (``ideal hydro''),
    and a reduced nucleon size (``smaller nucleon''). Error bars indicate statistical
    uncertainties from the finite event sample.} 
    \label{fig:pt_fluctuation_model_response}%
\end{figure}

We next examine how $R_{p_T}\!\equiv\!\sqrt{C_m}/\langle p_T\rangle$ responds to
variations of key model ingredients, focusing on Au+Au collisions at
$\sqrt{s_{NN}}=200$~GeV.  The results are summarized in
Fig.~\ref{fig:pt_fluctuation_model_response}, where the relative change of
$R_{p_T}$ is shown with respect to a common
baseline setup corresponding to the Bayesian-calibrated reference model~\cite{JETSCAPE:2020shq,JETSCAPE:2020mzn},
which includes shear and bulk viscosity, finite nucleon size ($w\!=\!1.12$ fm), and a
hadronic afterburner. To emphasize the intrinsic sensitivity of the observable,
all results are presented as fractional deviations from this baseline.

Two representative centrality classes, 30--40\% and 50--60\%, are shown.  This
choice allows us to contrast a mid-central regime, where collective dynamics
are well developed, with a more peripheral regime, where multiplicities are
smaller and different fluctuation sources compete more strongly.  As illustrated
in Fig.~\ref{fig:pt_fluctuation_model_response}, the response of $R_{p_T}$ to
individual model modifications is clearly centrality dependent and does not
follow a universal ordering across all variants.

Turning off bulk viscosity (``no bulk'') leads to a small change in
$R_{p_T}$, with a slightly negative response in 30--40\% collisions and
a positive response in the more peripheral 50--60\% class. This pattern
suggests that bulk-viscous effects contribute more visibly to
transverse-momentum fluctuations in dilute systems, where bulk-pressure
variations are less efficiently damped. However, the overall magnitude
of the response is modest in both centrality classes. Since turning off bulk viscosity hardens the single-particle spectra, it simultaneously enhances the correlated covariance sampled within the fixed $p_T$ window while increasing the mean transverse momentum in the denominator, leading to a partial cancellation in the normalized ratio $R_{p_T}$.

In contrast, the ideal-hydrodynamic limit (``ideal hydro''), in which
shear viscosity is also switched off and viscous damping is absent,
produces a systematic enhancement of \(R_{p_T}\) in both centrality
classes, with a larger relative effect in 50--60\% collisions. This
behavior is consistent with the interpretation of \(R_{p_T}\) as a
genuine fluctuation observable that is sensitive to dissipative
smoothing of event-by-event radial expansion. A comparison of the ``no
bulk'' and ``ideal hydro'' cases shows that \(R_{p_T}\) responds more
strongly to the removal of shear viscosity than to bulk viscosity within
the present setup, indicating that shear-driven damping of transverse
velocity gradients plays a dominant role in suppressing correlated
momentum fluctuations.

This sensitivity pattern contrasts with that of the
\(p_T\)-differential radial-flow fluctuation observable \(v_0(p_T)\),
which has been found to be more sensitive to bulk viscosity and
comparatively insensitive to shear effects
\cite{Parida:2024ckk,Du:2025dpu,ALICE:2025iud,ATLAS:2025ztg}. The
difference reflects the distinct physical information encoded in the two
observables. While \(R_{p_T}\) integrates momentum correlations over a
broad \(p_T\) range and is therefore primarily governed by the overall
coherence of event-by-event momentum shifts, making it particularly
susceptible to shear-induced smoothing, \(v_0(p_T)\) resolves the
momentum-dependent redistribution of radial-flow fluctuations and is
thus more directly influenced by bulk-viscous effects that modify the
spectral shape. 

Additional insight is provided by removing the hadronic afterburner (``no afterburner'') which produces opposite
trends in the two centrality bins: $R_{p_T}$ decreases in mid-central
collisions but increases in peripheral ones. This behavior reflects a
competition between correlation buildup during the hadronic stage and
multiplicity-driven dilution effects, whose relative importance changes
with system size.

Among the variations considered, reducing the nucleon size from
$w\!=\!1.12$~fm to $0.8$~fm (``smaller nucleon'') yields the largest and most
robust enhancement of $R_{p_T}$ across both centralities, underscoring the
strong sensitivity of the observable to initial-state granularity.
A smaller nucleon width increases event-by-event fluctuations of local pressure
gradients and hence amplifies fluctuations of the collective radial flow ($\beta_T$).
At a schematic level, the correlated momentum shift induced by radial expansion
can be expressed as
$\delta p_T \sim m_T\,\delta\beta_T$,
so that the two-particle covariance scales parametrically as
\begin{equation}
C_m \sim \langle \delta p_T\,\delta p_T \rangle
      \sim \langle m_T^2\rangle\,\langle \delta\beta_T^2\rangle .
\end{equation}
Reducing $w$ enhances $\langle \delta\beta_T^2\rangle$ by increasing the
granularity of the initial conditions, thereby strengthening collective
momentum correlations and leading to a larger $R_{p_T}$.

Taken together, Fig.~\ref{fig:pt_fluctuation_model_response}
demonstrates that $R_{p_T}$ is sensitive to multiple stages of the
collision dynamics, including initial-state structure, viscous
hydrodynamic evolution, and late-stage hadronic effects. The absence of
a universal response pattern across centrality underscores that the
observable probes competing fluctuation mechanisms whose relative
contributions evolve with system size. Rather than a drawback, this
feature makes $R_{p_T}$ a valuable complementary constraint on the
modeling of event-by-event dynamics, particularly when combined with
observables that primarily constrain average collective flow.

\subsection{Impact of kinematic acceptance on the beam-energy dependence}
\label{subsec:scaled_cut}
In Fig.~\ref{fig:pt_fluctuation_ptcut}, we presented a baseline comparison between model calculations and STAR data at $\sqrt{s_{NN}}=200$~GeV, using the experimental kinematic window $0.2<p_T<2$~GeV. As a consistency check, we also examined the sensitivity of the result to variations of the lower $p_T$ cut within the model and found that modest changes of $p_T^{\min}$ lead to visible shifts in the magnitude of the integrated fluctuation observable. This observation already indicates that, for $p_T$-integrated correlation
measures, the numerical value does not solely reflect the underlying
dynamics, but also depends on how the momentum-space correlation structure
is projected onto a finite kinematic acceptance. This projection effect
must therefore be carefully controlled when comparing results across
different beam energies.

This motivates a more systematic examination of kinematic effects when
comparing results across different beam energies. In experimental
measurements, the commonly used acceptance windows are not identical:
at RHIC, STAR typically employs $0.2<p_T<2$~GeV at
$\sqrt{s_{NN}}=200$~GeV \cite{STAR:2019dow}, while at the LHC, ALICE uses
$0.15<p_T<2$~GeV at $\sqrt{s_{NN}}=2.76$ and $5.02$~TeV \cite{ALICE:2014gvd,ALICE:2024apz}. However, even if
identical absolute $p_T$ cuts were applied at all energies, the
corresponding momentum window would not probe the same relative portion
of the underlying particle spectrum. As the characteristic transverse
momentum scale of the system increases with beam energy, and varies with
centrality at fixed energy, the same fixed $p_T$ range samples different
regions of the soft momentum distribution and therefore different
components of the underlying momentum correlations.

The physical origin of this effect can be understood from the structure
of the single-particle transverse-momentum spectra themselves.
For a fixed centrality class, we find that identified-particle $p_T$
spectra at different beam energies approximately collapse onto a common
curve when expressed in terms of the dimensionless variable
$p_T/\langle p_T\rangle$, up to an overall normalization factor. This
approximate scaling indicates that the ensemble-averaged mean transverse
momentum $\langle p_T\rangle$ provides a natural soft momentum scale of
the system. Consequently, a fixed absolute $p_T$ window corresponds to
different relative locations in scaled momentum space as the beam energy
or centrality changes, even when the same nominal cuts are applied.

\begin{figure}[t]
    \centering 
    \includegraphics[width= 0.8\linewidth]{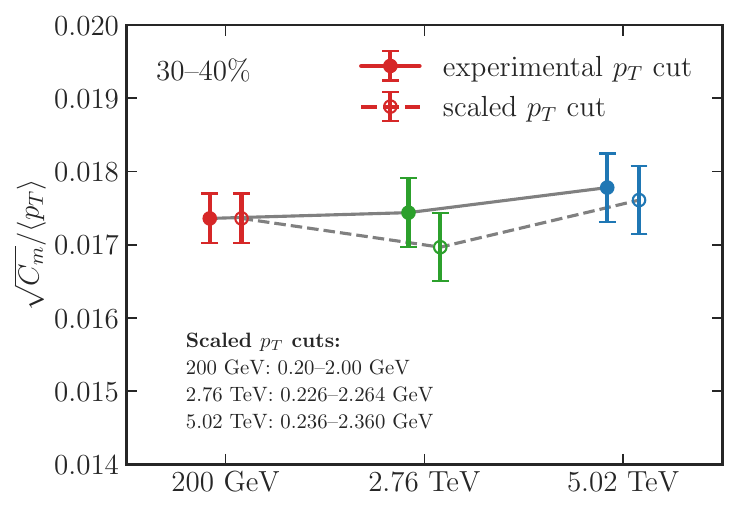}
    \caption{Beam-energy dependence of $\sqrt{C_m}/\langle p_T\rangle$ for charged hadrons in 30--40\% central
    collisions, comparing results obtained with experimental
    kinematic cuts (solid symbols) and with scaled $p_T$ cuts (open symbols).
    The experimental cut setup follows the published experimental acceptances:
    $0.2<p_T<2$~GeV at $\sqrt{s_{NN}}=200$~GeV (STAR) and
    $0.15<p_T<2$~GeV at $\sqrt{s_{NN}}=2.76$ and $5.02$~TeV (ALICE).
    The scaled-cut results use energy-dependent windows defined relative to
    the charged-particle mean transverse momentum and anchored to the
    STAR acceptance at 200~GeV, yielding
    $0.226<p_T<2.264$~GeV at 2.76~TeV and
    $0.236<p_T<2.360$~GeV at 5.02~TeV.} 
    \label{fig:scaled_cut_energy}%
\end{figure}

Guided by this observation, we introduce a ``scaled-cut'' prescription
designed to align the kinematic acceptance relative to a common soft
momentum scale when comparing different beam energies. To disentangle
genuine beam-energy systematics from kinematic projection effects,
Fig.~\ref{fig:scaled_cut_energy} compares two setups for mid-central
(30--40\%) collisions. The first corresponds to the ``experimental cut''
setup, in which the published acceptance windows are used at each beam
energy. The second employs a scaled acceptance defined as
\begin{align}
p_T^{\min}(\sqrt{s_{NN}}) &= \alpha\,\langle p_T\rangle_{\rm ch}(\sqrt{s_{NN}}),\\
p_T^{\max}(\sqrt{s_{NN}}) &= \beta\,\langle p_T\rangle_{\rm ch}(\sqrt{s_{NN}}),
\end{align}
where $\langle p_T\rangle_{\rm ch}$ is the charged-particle mean
transverse momentum in the same centrality class. The constants
$\alpha$ and $\beta$ are fixed by anchoring the scaled window to the STAR
acceptance at $\sqrt{s_{NN}}=200$~GeV, $0.2<p_T<2$~GeV. This procedure
yields energy-dependent momentum windows, $0.226<p_T<2.264$~GeV at 2.76~TeV and
    $0.236<p_T<2.360$~GeV at 5.02~TeV, that shift both the lower and
upper edges upward at LHC energies, reflecting the harder underlying
spectra.

Figure~\ref{fig:scaled_cut_energy} shows that, under the experimental cuts,
the integrated fluctuation measure exhibits an increase with beam energy.
In contrast, when the scaled-cut prescription is applied, the results at
both LHC energies are reduced relative to the fixed-cut case, with a
larger suppression observed at $\sqrt{s_{NN}}=2.76$~TeV. This behavior
naturally follows from the shift of the scaled acceptance toward higher
transverse momenta, which samples a harder region of the momentum
distribution than the published experimental cuts.

The reduction of $\sqrt{C_m}/\langle p_T\rangle$ under the scaled-cut
prescription is closely connected to the $p_T$-cut dependence discussed
in Sec.~\ref{subsec:baseline_200}. There, lowering the minimum $p_T$ cut
enhanced the fluctuation signal by including a larger fraction of very
soft particles, where collective radial dynamics are strongest. In the
present case, the scaled cuts act in the opposite direction at LHC
energies by increasing both $p_T^{\min}$ and $p_T^{\max}$ relative to the
experimental acceptance, thereby suppressing contributions from the
softest, most strongly correlated part of the spectrum.

Within current uncertainties, the resulting beam-energy dependence under
scaled cuts may even appear non-monotonic. We emphasize, however, that
this behavior should not be over-interpreted as evidence for a genuine
non-monotonic dynamical trend. Rather, it reflects a general kinematic
projection mechanism inherent to $p_T$-integrated fluctuation observables.
Because $R_{p_T}$ is dominated by correlations carried by the soft part of
the spectrum, acceptance windows that shift toward higher relative
momenta systematically suppress the measured fluctuation magnitude by
reducing the weight of the most strongly correlated low-$p_T$ particles.
Scaled cuts make this suppression explicit by construction, while fixed
absolute $p_T$ cuts can induce it implicitly as the characteristic
momentum scale of the system changes.

As a consequence, part of the observed beam-energy dependence of
$p_T$-integrated fluctuation measures  \cite{STAR:2019dow} may arise from kinematic projection
effects rather than changes in the underlying correlation strength
itself (see~\ref{app:kinematic_projection}). When fixed absolute $p_T$ cuts are used across energies, the
acceptance effectively drifts relative to the soft momentum scale of the
system, altering the balance between soft and semi-soft contributions as
the spectra harden or soften with $\sqrt{s_{NN}}$. The scaled-cut
comparison presented here provides a controlled demonstration of this
effect at top RHIC and LHC energies and highlights its direct relevance
for RHIC Beam Energy Scan measurements~\cite{STAR:2008med,STAR:2017sal,Bzdak:2019pkr,Du:2023efk,Du:2024wjm}, where the progressive softening of the
single-particle spectra can significantly enhance acceptance-induced
biases in the apparent beam-energy systematics.

\subsection{Species dependence and effective kinematic projection}
\label{subsec:species_dependence}

Identified-particle measurements provide a more differential probe of
transverse-momentum fluctuations by isolating their dependence on hadron
mass and species composition. They offer a direct test of whether the
observed correlations exhibit the mass dependence expected from
fluctuating radial flow and help clarify how changes in particle
composition may influence inclusive fluctuation measurements. At
present, such species-resolved measurements of transverse-momentum
fluctuations are not available experimentally, and the results presented
here should therefore be viewed as exploratory predictions within the
calibrated framework.

\begin{figure}[t]
    \centering 
    \includegraphics[width= 0.8\linewidth]{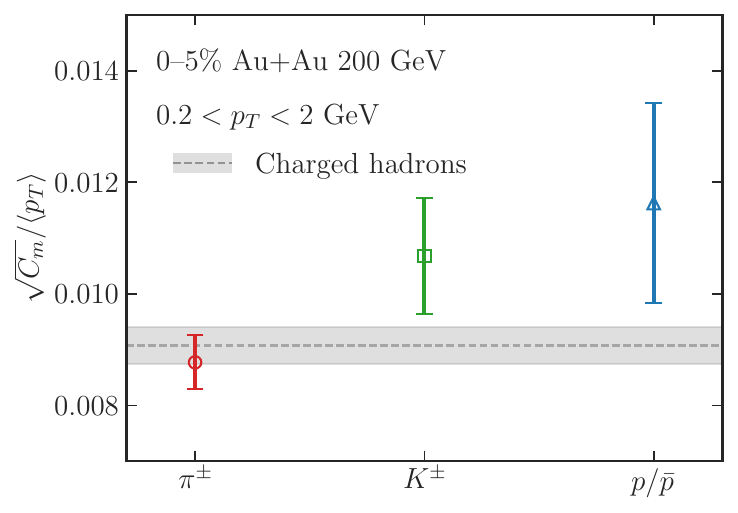}
    \caption{Species dependence of 
    $\sqrt{C_m}/\langle p_T\rangle$ for identified $\pi^{\pm}$, $K^{\pm}$, and
    $p/\bar p$ in central (0--5\%) Au+Au collisions at
    $\sqrt{s_{NN}}=200$~GeV, evaluated in a common kinematic window
    $0.2<p_T<2$~GeV.
    The shaded band and dashed line indicate the corresponding result for
    inclusive charged hadrons in the same centrality and acceptance.} 
    \label{fig:species_dependence}%
\end{figure}

Figure~\ref{fig:species_dependence} shows the fluctuation observable
$\sqrt{C_m}/\langle p_T\rangle$ evaluated separately for identified
$\pi^{\pm}$, $K^{\pm}$, and $p/\bar p$ in central (0--5\%) Au+Au
collisions at $\sqrt{s_{NN}}\!=\!200$~GeV, using a common kinematic window
$0.2<p_T<2$~GeV for all species. For reference, the corresponding result
for inclusive charged hadrons in the same centrality and acceptance is
also indicated.
A clear mass ordering is observed: the fluctuation measure increases
systematically from pions to kaons to protons. Since the mean transverse
momentum itself grows with particle mass,
$\langle p_T\rangle_p > \langle p_T\rangle_K > \langle p_T\rangle_\pi$,
this ordering implies an even stronger mass dependence of the absolute
covariance $\sqrt{C_m}$. In other words, heavier particles not only have
larger mean momenta, but also exhibit substantially stronger
event-by-event momentum correlations.

This behavior is naturally understood as a consequence of fluctuating
radial flow. Event-wise variations of the collective transverse velocity
field lead to momentum shifts, $\delta p_T \sim m_T\,\delta\beta_T$, that scale approximately with
particle mass, such that heavier hadrons experience larger absolute
responses to the same underlying flow fluctuation. As a result, the
two-particle covariance $C_m$, which encodes correlated momentum
fluctuations, is amplified for heavier species. The observed mass
ordering therefore provides further evidence that the dominant
contribution to the fluctuation observable originates from collective
dynamics rather than from few-particle or non-collective sources.

Importantly, the species dependence observed in
Fig.~\ref{fig:species_dependence} is closely connected to the kinematic
acceptance effects discussed in
Secs.~\ref{subsec:baseline_200} and~\ref{subsec:scaled_cut}. Although the
same absolute $p_T$ window is applied to all species, this window
corresponds to different regions of the underlying spectra when measured
relative to each species’ characteristic momentum scale. Expressed in
units of $p_T/\langle p_T\rangle$, the fixed window probes a softer and
more flow-dominated region for protons than for pions, while sampling a
comparatively harder region for lighter species. As a result, the fixed
$p_T$ cut does not merely reveal the intrinsic mass dependence of the
fluctuations but systematically amplifies it by enhancing the relative
weight of soft, strongly correlated particles for heavier hadrons.
Consequently, different particle species effectively project different
parts of the underlying momentum-correlation structure even under
identical kinematic cuts.

From this perspective, the mass ordering seen here represents an
internal analogue of the explicit $p_T$-cut dependence discussed earlier.
Varying the particle species at fixed acceptance shifts the effective
location of the kinematic window in scaled momentum space in much the
same way as changing the lower or upper $p_T$ cut for a single species.
The enhanced sensitivity of heavier hadrons to both collective flow
fluctuations and kinematic acceptance effects further reinforces the
interpretation that the measured transverse-momentum correlations are
dominated by global, event-wise dynamics of the expanding medium.

Within this context, the present species-resolved results for light hadrons
provide a controlled baseline for collective dynamics at kinetic freeze-out.
Heavy-flavor particles offer valuable complementary information on the degree of
thermalization and coupling between heavy quarks and the medium, and fluctuation
observables could, in principle, provide additional sensitivity beyond average
flow measurements; indeed, within the JETSCAPE framework, heavy-flavor dynamics---particularly for open heavy mesons---have been studied using dedicated transport
modules \cite{JETSCAPE:2022hcb, JETSCAPE:2023ikg}, with significant progress in describing heavy-flavor $R_{AA}$. A systematic extension of the present fluctuation analysis to heavy-flavor
hadrons would therefore require a dedicated treatment of non-equilibrium
production and transport effects and is beyond the scope of this work.

\section{Conclusions}
\label{sec:conclusions}
In this work, we have presented a systematic model study of event-by-event
transverse-momentum fluctuations in relativistic heavy-ion collisions,
focusing on the observable
$R_{p_T}\equiv\sqrt{C_m}/\langle p_T\rangle$
and its sensitivity to soft-sector dynamics and kinematic acceptance.
Using a Bayesian-calibrated event-by-event hydrodynamic framework, we
performed baseline comparisons to RHIC data, investigated controlled
variations of model ingredients, and examined the interplay between
dynamical and kinematic effects in beam-energy systematics.

At $\sqrt{s_{NN}}=200$~GeV, the model reproduces the quantitative centrality
dependence of the STAR measurements within the experimental acceptance.
The remaining discrepancy in magnitude is traced to a pronounced
sensitivity of $R_{p_T}$ to the very soft part of the transverse-momentum
spectrum, indicating that the integrated fluctuation signal is dominated
by long-wavelength collective dynamics carried by low-$p_T$ particles.
This observation highlights the importance of kinematic acceptance in
the interpretation of $p_T$-integrated fluctuation observables.

By varying individual model ingredients at fixed beam energy, we showed
that $R_{p_T}$ responds to multiple stages of the collision evolution,
including initial-state granularity, viscous hydrodynamic damping, and
late-stage hadronic rescattering. The magnitude and even the sign of these
responses depend on centrality, underscoring that transverse-momentum
fluctuations probe competing fluctuation mechanisms whose relative
importance evolves with system size. This behavior demonstrates that
$R_{p_T}$ provides information complementary to average flow observables
and offers an independent handle on event-by-event collective dynamics.

When comparing model results across beam energies, we demonstrated that a
substantial part of the apparent energy dependence of $R_{p_T}$ obtained
under fixed experimental $p_T$ cuts can arise from kinematic projection
effects. Introducing a scaled-cut prescription that aligns the acceptance
relative to the characteristic soft momentum scale qualitatively modifies
the observed energy dependence, emphasizing the need for acceptance-aware
comparisons in RHIC-to-LHC studies and beam energy scan programs.

Finally, we explored the species dependence of transverse-momentum
fluctuations in central Au+Au collisions at $\sqrt{s_{NN}}=200$~GeV.
A clear mass ordering is observed, with heavier hadrons exhibiting
stronger momentum correlations. This behavior is naturally explained by
fluctuating radial flow and provides an internal analogue of the
$p_T$-cut dependence discussed earlier, illustrating how different
particle species effectively probe different regions of the underlying
momentum-correlation structure even within identical kinematic windows.

Taken together, our results establish transverse-momentum fluctuation
observables as sensitive probes of collective dynamics whose quantitative
interpretation requires careful control of kinematic acceptance and
projection effects. Future experimental measurements that explore
alternative $p_T$ windows, identified-particle fluctuations, and
differential extensions of the present analysis will provide valuable
opportunities to further isolate dynamical contributions and to sharpen
the sensitivity of $p_T$-correlation observables to the properties of the
expanding quark--gluon plasma, in a manner complementary to traditional
flow-based constraints.

\section*{Acknowledgements}

The author acknowledges helpful discussions with workshop participants,
especially I.~Karpenko, R.~Manikandhan and T.~Reichert, at the workshop
\emph{The QCD Critical Point: Are We There Yet?} hosted by the Institute for
Nuclear Theory (INT) at the University of Washington, which stimulated
the present study. The author thanks the INT for its warm hospitality and
stimulating research environment.
This work was supported in part by the U.S. Department of Energy through
the INT under Grant No.~DE-FG02-00ER41132, and in part by the U.S.
Department of Energy, Office of Science, Office of Nuclear Physics under
Grant No.~DE-AC02-05CH11231. Computational resources were provided by the
Ohio Supercomputer Center~\cite{OhioSupercomputerCenter1987}.
The author acknowledges the use of ChatGPT for assistance with grammar
refinement and clarity improvement during manuscript preparation.

\appendix
\section{Relation between the STAR and ALICE transverse-momentum fluctuation definitions}
\label{app:method_comparison}

In Secs.~\ref{subsec:observable_definition} and
\ref{subsec:experimental_definitions}, we introduced the transverse-momentum
fluctuation observable
$R_{p_T}=\sqrt{C_m}/\langle p_T\rangle$ and summarized the STAR and ALICE
experimental prescriptions in a unified framework.
In this appendix, we provide a compact analytical relation between the two
definitions, clarifying why they yield nearly identical results in the present
analysis.

For a given event $k$ with multiplicity $N_k$, we denote the event-wise mean
transverse momentum by
$[p_T]_k=\frac{1}{N_k}\sum_i p_{T,i}$ and the ensemble-averaged mean by
$\langle p_T\rangle=\langle [p_T]_k\rangle$.
The STAR and ALICE definitions differ only in the choice of reference mean
$p_{T,\mathrm{ref}}$ entering the event-wise covariance
(Eq.~\eqref{eq:Cm_event}):
\[
p_{T,\mathrm{ref}}^{\mathrm{STAR}}=\langle p_T\rangle,
\qquad
p_{T,\mathrm{ref}}^{\mathrm{ALICE}}=[p_T]_k.
\]

Introducing the decomposition
$p_{T,i}-\langle p_T\rangle=(p_{T,i}-[p_T]_k)+([p_T]_k-\langle p_T\rangle)$,
the STAR and ALICE single-event covariances,
$C_k^{\mathrm{STAR}}$ and $C_k^{\mathrm{ALICE}}$, are related by the exact identity
\begin{equation}
C_k^{\mathrm{STAR}}
=
C_k^{\mathrm{ALICE}}
+
\big([p_T]_k-\langle p_T\rangle\big)^2 .
\label{eq:star_alice_relation}
\end{equation}
Here $C_k$ denotes the event-wise covariance defined in
Eq.~\eqref{eq:Cm_event}, prior to the averaging over events that yields
the experimentally reported quantity $C_m$.
Averaging over events within a fixed centrality class yields
\begin{equation}
\langle C_k^{\mathrm{STAR}}\rangle
=
\langle C_k^{\mathrm{ALICE}}\rangle
+
\mathrm{Var}([p_T]_k),
\label{eq:star_alice_relation_avg}
\end{equation}
where $\mathrm{Var}([p_T]_k)\equiv\big\langle\big([p_T]_k-\langle p_T\rangle\big)^2\big\rangle$ is the event-by-event variance of the mean
transverse momentum.

Equation~\eqref{eq:star_alice_relation_avg} shows that the STAR definition contains,
in addition to the ALICE-style two-particle covariance, an explicit contribution
from fluctuations of the event-wise mean $[p_T]$ within the centrality bin.
As discussed in Sec.~\ref{subsec:experimental_definitions}, the magnitude of this
additional term depends on the particle multiplicity and the width of the
centrality selection.

In the present analysis, the use of relatively narrow centrality bins and the
large multiplicities characteristic of RHIC and LHC collisions render
$\mathrm{Var}([p_T]_k)$ numerically small compared to the genuine two-particle
correlation contribution.
Consequently, the STAR and ALICE prescriptions yield nearly identical values of
$R_{p_T}$ when applied to the same set of model events with identical kinematic
cuts, as explicitly demonstrated in Sec.~\ref{subsec:baseline_200}.
This near-equivalence justifies our use of a single representative definition in
the presentation of results, while retaining both prescriptions as an internal
consistency check on the robustness of the extracted fluctuation signal. However, in very peripheral collisions or for wide centrality selections, the variance term may become non-negligible.

\section{Kinematic projection effects and beam-energy systematics at RHIC BES}
\label{app:kinematic_projection}

The interpretation of the beam-energy dependence of the transverse-momentum
fluctuation observable
$R_{p_T}=\sqrt{C_m}/\langle p_T\rangle$
requires careful separation of genuine dynamical effects from kinematic
projection effects associated with the finite $p_T$ acceptance.
This appendix provides a unified explanation of how fixed absolute $p_T$ cuts
affect both the magnitude and the apparent energy dependence of $R_{p_T}$ at
RHIC BES energies, and why a substantial part of the observed
trend can arise from kinematics alone.

The two-particle covariance
\begin{equation}
C_m \;\sim\;
\int dp_T\,dp_T'
\left(\frac{dN}{dp_T}\right)
\left(\frac{dN}{dp_T'}\right)
\langle \delta p_T\,\delta p_T' \rangle
\end{equation}
is dominated by correlated momentum shifts generated by event-by-event
fluctuations of collective radial expansion.
To leading order, a fluctuation of the transverse flow velocity
$\delta\beta_T$ induces a correlated shift of single-particle momenta,
$\delta p_T \sim m_T\,\delta\beta_T$.
Although higher-$p_T$ particles experience larger absolute momentum shifts,
the rapidly falling single-particle yield, $dN/dp_T$, suppresses their statistical weight.
As a result, the dominant contribution to $C_m$ arises from the low-$p_T$
region, where particle production is largest.
This feature underlies the strong sensitivity of $C_m$ to the soft part of the
spectrum.

At a fixed beam energy and centrality, modifying the $p_T$ acceptance therefore
has a direct and predictable effect on $R_{p_T}$.
Extending the acceptance toward lower $p_T$ includes additional particles from
this correlation-dominated region, increasing the correlated covariance
$\sqrt{C_m}$.
At the same time, the ensemble-averaged mean transverse momentum
$\langle p_T\rangle$ decreases due to the enhanced statistical weight of soft
particles.
Both effects act coherently to increase $R_{p_T}$.
Conversely, restricting the acceptance to higher $p_T$ excludes soft correlated
pairs, reduces $\sqrt{C_m}$, and increases $\langle p_T\rangle$, leading to a
suppression of $R_{p_T}$.
This behavior reflects a purely kinematic projection effect and does not imply
any change in the underlying dynamics.

When a fixed absolute $p_T$ window is applied across different beam energies in
the RHIC BES, an additional effect enters.
As the beam energy is lowered, the characteristic momentum scale of the system
decreases, and the same fixed $p_T$ window probes progressively harder regions
of the spectrum when expressed in scaled units such as $p_T/\langle p_T\rangle$.
This kinematic drift systematically removes the lowest-$p_T$ particles that
carry the largest share of the correlated covariance.
As a result, $\sqrt{C_m}$ decreases with decreasing beam energy even if the
intrinsic strength of collective fluctuations were unchanged.

The denominator $\langle p_T\rangle$ is affected in the opposite direction:
excluding soft particles increases the measured mean transverse momentum.
This increase acts in the same direction as the suppression of $\sqrt{C_m}$, further reducing $R_{p_T}$; however, its quantitative impact is subleading compared to the loss of correlated soft pairs in the covariance.
The correlated covariance $C_m$ is a two-particle quantity dominated by the
multiplicity-rich soft sector, whereas $\langle p_T\rangle$ is a single-particle
moment that varies more smoothly with acceptance.
Consequently, the suppression of $\sqrt{C_m}$ dominates, and the net effect is a
decrease of $R_{p_T}$ as the beam energy is lowered.

Across the wide energy range of the RHIC BES, this kinematic drift becomes
pronounced.
The fixed absolute $p_T$ cut moves rapidly toward the tail of the spectrum at
lower energies, leading to a substantial loss of soft, collectively correlated
particles.
The resulting kinematic suppression of $R_{p_T}$ can therefore amplify the
apparent beam-energy dependence observed in experimental measurements.
Importantly, this mechanism operates independently of any genuine change in the
microscopic fluctuation dynamics and must be accounted for when interpreting
energy-dependent trends of $p_T$-integrated fluctuation observables.

Taken together, these considerations demonstrate that the decrease of
$R_{p_T}$ toward lower RHIC BES energies obtained with fixed absolute $p_T$ cuts
contains a significant kinematic component.
A robust interpretation of beam-energy systematics therefore requires careful
control of acceptance effects, either through scaled-cut prescriptions or
through complementary differential measurements that explicitly resolve the
momentum dependence of the underlying correlations.

\bibliographystyle{elsarticle-num} 
\bibliography{refs}

\end{document}